\documentclass[prd,showpacs,nofootinbib,twocolumn]{revtex4}
\usepackage{amsmath}
\usepackage{latexsym}
\usepackage{graphicx}
\usepackage{hyperref}

\newcommand{\beq}{\begin{equation}}
\newcommand{\eeq}{\end{equation}}
\newcommand{\bea}{\begin{eqnarray}}
\newcommand{\eea}{\end{eqnarray}}
\newcommand{\ba}{\begin{array}}
\newcommand{\ea}{\end{array}}
\newcommand{\Msun}{{\rm M}_{\odot}}

\newcommand{\eg}{e.~g.~}
\def\leq{\,\raise 0.4ex\hbox{$<$}\kern -0.8em\lower 0.62ex\hbox{$-$}\,}
\def\geq{\,\raise 0.4ex\hbox{$>$}\kern -0.8em\lower 0.62ex\hbox{$-$}\,}
\def\pm{\,\raise 0.4ex\hbox{$+$}\kern -0.8em\lower 0.62ex\hbox{$-$}\,}
\def\lsim{\,\raise 0.4ex\hbox{$<$}\kern -0.8em\lower 0.62ex\hbox{$\sim$}\,}
\def\gsim{\,\raise 0.4ex\hbox{$>$}\kern -0.8em\lower 0.62ex\hbox{$\sim$}\,}

\begin{document}

\title{The Status of Black-Hole Binary Merger Simulations with Numerical Relativity}

\author{Sean T. McWilliams}
\email{sean@astro.columbia.edu}
\affiliation{Institute for Strings, Cosmology and Astroparticle Physics (ISCAP), Columbia University, New York, NY 10027}
\affiliation{Physics Department, Princeton University, Princeton, NJ 08544}

\date{\today}

\begin{abstract}
The advent of long-term stability in numerical relativity has yielded a windfall of answers to long-standing questions regarding the dynamics
of space-time, matter, and electromagnetic fields in the strong-field regime of black-hole binary mergers.  In this review, we will briefly summarize
the methodology currently applied to these problems, emphasizing the most recent advancements.  We will discuss recent results of
astrophysical relevance, and present
some novel interpretation.  Though we primarily present a review, we also present a simple analytical model for the 
time-dependent Poynting flux from two orbiting
black holes immersed in a magnetic field, which compares favorably with recent numerical results.
Finally, we will discuss recent advancements in our theoretical understanding of merger dynamics and gravitational
waveforms that have resulted from interpreting the ever-growing body of numerical relativity results.
\end{abstract}


\maketitle

\section{Introduction}

In the last five years, the field of gravitational wave (GW) astronomy has witnessed a remarkable convergence of experimental and theoretical
achievement.  The current generation of GW interferometers, particularly the Laser Interferometer Gravitational Wave
Observatory (LIGO) and VIRGO detectors, has positioned this community on the cusp of discovery, by achieving
their design sensitivities and making the first direct detection of GWs a real possibility.  Meanwhile, the GW
signature from the merger of a black-hole binary (BHB), expected to be the most common source for these and future observatories,
went from being a complete unknown to a reasonably well-understood, surprisingly smooth transition between the earlier inspiral and the final ringdown,
thanks to tremendous advancements in the field of numerical relativity (NR).
In this article, we will limit our scope to the
advancements that have occurred in just the past year, with only a minimal amount of background information,
in order to provide a snapshot of the most current research in the field.
For a review of progress in the preceding year, we refer the reader to \cite{Hinder:2010vn}, and for a broader review of NR
achievements, please see \cite{Hannam:2009rd,Centrella:2010mx}.  We consider it an indicator of the healthy progress of the field that,
in addition to the general review articles typical of any field,
in recent years the field of NR has progressed rapidly enough to warrant an annual review of the past year's progress.

In Sec.~\ref{sec:background}, we give a brief review of the equations being solved by numerical relativists, and the methods that are most typically
applied to solve them. 
In Sec.~\ref{sec:res}, we discuss the most recent achievements in NR methodology, and the simulations of
numerically challenging systems of astrophysical interest that have been facilitated.
In Sec.~\ref{sec:theory}, we briefly mention novel theoretical developments resulting from the interpretation of numerical results.
We summarize and conclude in Sec.~\ref{sec:conc}.

\section{Standard Methodology}
\label{sec:background}

The community of numerical relativists attempts to solve Einstein's equations in the strong-field regime numerically (hence the name).  During the early,
approximately-adiabatic inspiral, we can treat the system perturbatively through expansions in $v/c$ or other small parameters related to it.
After the merger, we can again treat the system perturbatively, as the quasinormal ringing of small deviations from a Kerr background.
However, during the merger, unless the masses are disparate and $q=m_1/m_2$ provides a tractably small expansion parameter, we can no longer
treat the system perturbatively, and are forced to solve the full equations, given by
\beq
R_{\mu\nu} - \frac{1}{2} Rg_{\mu\nu} = 8\pi T_{\mu\nu} \,,
\label{eq:einstein}
\eeq
where $R_{\mu \nu}$, $R$, $g_{\mu \nu}$, and $T_{\mu\nu}$ are the Ricci tensor, its trace, and the metric and stress-energy tensors, respectively.

In the case of BHBs in vacuum, $T_{\mu\nu}$ vanishes.  The resulting equation is generally evolved by splitting up
space-time into a sequence of three-dimensional hypersurfaces.  The choice of how to make this split is not unique, as different
evolution equations with different
variables can be used, as well as different coordinates.  The coordinate freedom is fixed through the choice of dividing space-time into hypersurfaces.
The lapse, $\alpha$, determines the coordinate time interval between one hypersurface and the next.  The shift, $\beta_i$, determines the
motion of the three spatial coordinates within each hypersurface.
The splitting of space-time results in a set of evolution equations, plus constraints
that need to be satisfied.  To the extent that the constraints vanish, all evolution equations must be analytically equivalent.
However, different choices will have different numerical properties.  Indeed, the four decades that passed between the first attempts
to perform NR simulations and the recent breakthroughs were spent, in large part, searching for formulations that would be numerically stable. 

At present, there are primarily three analytic forms for Einstein's equations that are actively being applied to the BHB problem: the BSSN formulation
\cite{Nakamura:1987zz, Shibata:1995we,Baumgarte:1998te} and two different Generalized Harmonic formulations 
\cite{Pretorius:2004jg, Pretorius:2005gq, Lindblom:2005qh}.  Most typically, the equations are spatially discretized with finite differences (with
the exception of {\sc SpEC} \cite{Scheel:2006gg}, which uses pseudo-spectral methods).  The state-of-the-art codes employ
eighth-order spatial differencing, but most codes have a mixture of orders for different parts of the code.  All implementations of which we are aware
discretize in time using Runge-Kutta methods, usually to fourth-order accuracy in the time step.  Given the disparate scales involved in the
problem ($r \gsim 100M$ for the wave zone, $\Delta \lsim M/100$ for the near-field), a range of resolutions within each
simulation is required, and this is achieved
in a number of different ways by different groups.  There are many codes being actively used and developed
\cite{Baiotti:2010zf}, \cite{Imbiriba:2004tp}, \cite{Pretorius:2005gq}, \cite{Campanelli:2005dd},
\cite{Sperhake:2006cy}, \cite{Scheel:2006gg}, \cite{Pollney:2009yz}, \cite{Giacomazzo:2007ti}, \cite{Yamamoto:sacra}, 
and comparisons have been published that verify the consistency of results among different groups
\cite{Baker:2007fb,Hannam:2009hh}.

In the presence of matter and/or electromagnetic (EM) fields, the equations
governing the evolution of space-time must be coupled to an appropriate stress-energy tensor.  It is noteworthy that matter is not needed for
$T_{\mu \nu}$ to be nonvanishing.  In the case of the Einstein-Maxwell system (see \eg \cite{Mosta:2009rr}), the sources for EM
fields are not included in the stress-energy tensor.  The inclusion of these sources, the fields themselves, and the equations of hydrodynamics,
all coupled to Einstein's equations, constitute the equations of general relativistic magneto-hydrodynamics (GRMHD),
while the coupling of the hydrodynamics and Einstein's equations, without EM fields, 
make up the general relativistic hydrodynamics (GRHD) equations.  The recent advent
of GR(M)HD codes has yielded some of the most interesting science in NR, and perhaps in all of theoretical astrophysics, in the
past year, and will be discussed in Sec.~\ref{sec:grmhd}.

\section{Novel Methodology and Astrophysics}
\label{sec:res}

In this section, we will discuss novel developments in NR methodology, and the novel astrophysical results that those
developments have facilitated.

\subsection{``Extreme'' mass ratios} 

Whereas the typical definition of extreme mass ratio systems is $q\sim 10^{-6}$, such systems are of little interest to numerical relativists,
as they can be treated well with perturbative methods, and they are, at present, completely impossible to simulate non-perturbatively.  However,
for more intermediate mass ratios, $q\geq 1/100$, it is now possible to perform short simulations which include 1--2 orbits of the late inspiral,
the merger, and the ringdown.  This is an impressive achievement, because the resolution demands are set by the smaller black hole, so
that a black hole 100 times smaller than its companion requires 100 times finer resolution.  The trailblazers in the endeavor
to simulate these systems have been the RIT group, using their {\sc LazEv} code \cite{Zlochower:2005bj}.  Their first effort involved combining a 
full numerical evolution of the punctures with a perturbative treatment of the radiation \cite{Lousto:2010tb}.  Later, they validated this approach
by comparing with fully numerical simulations for a $q=1/10$ system \cite{Lousto:2010qx}.  Most recently, they have achieved the aforementioned
$1/100$ mass ratio through a combination of improved gauge conditions, an improved mesh refinement scheme, and an improved allotment of
cpu cycles \cite{Lousto:2010ut}.

\subsection{Toward Maximal Kerr Initial Data}

The Caltech-Cornell-CITA collaboration has overcome the obstacles of generically stable black-hole binary evolutions, and have demonstrated
the ability to stably evolve systems with moderate mass ratios $q\leq 1/2$ \cite{Szilagyi:2009qz} and extremely high spins
$a=S/M^2\leq 0.95$ \cite{Lovelace:2010up}, including merger and ringdown, using their 
{\sc SpEC} code \cite{Scheel:2006gg}.  To facilitate the simulation of stable near-extremal spins, they had to employ novel methodology
in order to exceed the bound of $a=S/M^2\lsim 0.93$ that is an intrinsic characteristic of Bowen-York initial data \cite{Bowen:1980yu}.  
More specifically,
while the spin in Bowen-York initial data can be set arbitrarily high, it will quickly relax to the aforementioned bound, so that larger
spins cannot be evolved with that approach.  To this end, the authors of \cite{Lovelace:2010up} applied
a novel method \cite{Lovelace:2008tw} wherein the extended-conformal-thin-sandwich equations \cite{York:1998hy} were combined with a
conformally-curved metric, constructed through the weighted superposition of the metrics for two boosted, spinning black holes in Kerr-Schild
coordinates.  Using this approach facilitated the simulation of a pair of equal-mass black holes with spins of $a=0.95$ anti-aligned
with the orbital angular momentum (thereby minimizing the total system angular momentum) \cite{Lovelace:2010up}.  We note that the Illinois
group also developed a novel approach for near-extremal initial data \cite{Liu:2009}, but it has not yet been demonstrated in a binary evolution.

\subsection{Nonlinear memory}

An interesting consequence of the nonlinear nature of GW propagation is the emission of GWs generated
by GWs.  This effect has long been established analytically, and is often referred to as the ``Christodoulou memory''
\cite{Christodoulou:mem}.  However, this effect is exceedingly small, and is therefore difficult to simulate, as systematic error in the
simulations must be smaller than the effect.  The GW memory, though small, could conceivably be detectable by the
space-based Laser Interferometer Space Antenna (LISA) \cite{Favata:2008yd}.  The authors of \cite{Pollney:mem} were able to accurately calculate
the memory effect from the final merger and ringdown, by using novel numerical methodology implemented in their {\sc Llama} code, which includes
multipatch capabilities \cite{Pollney:2009ut,Pollney:2009yz} and Cauchy characteristic extraction (CCE) \cite{Reisswig:prl,Reisswig:cce}.
The multipatch code facilitates multiple coordinate patches in different regions, allowing traditional Cartesian grids around the near zone,
with a smooth transition to a six-patch polar grid in the wave zone.  Since the waves do not gain angular structure in the wavezone, this allows one
to increase the grid size by only adding radial elements, thereby greatly decreasing the cost of large grids, and permitting much more accuracy
in the region where the waves propagate.  The implementation of CCE represents the first fully gauge invariant GW calculation,
wherein the GWs are formally measured at infinity, rather than extrapolating to infinity from a sequence of finite extraction radii 
(though we note that progress has also been made in the last year on compactification, 
including infinity explicitly in the computational domain \cite{Zeng}).
Using {\sc Llama}, the authors of \cite{Pollney:mem} were able to accurately resolve the nonlinear
memory for the first time.  We note that, in \cite{Favata:2008yd,Pollney:mem}, the $h_{20}$ component of the GW strain
(which contains the memory) appears to
be a significant fraction of the fundamental quadrupole modes, so that some readers might be confused why the signal-to-noise ratio (SNR) of
the memory contribution is roughly two orders of magnitude below the SNR of the total signal (see Fig. 2 of \cite{Favata:2008yd}).  In this context,
one needs to bear in mind that constant offsets in strain are gauge-dependent and undetectable.  Therefore, only the derivative of the
memory contribution to the GW strain contributes to the SNR \cite{Bakercomm}.

\subsection{GR(M)HD}
\label{sec:grmhd}

Perhaps the most exciting developments have been the coming-of-age of GR(M)HD codes.  There are currently four codes capable of GRMHD
with dynamical space-time: {\sc WhiskyMHD} \cite{Giacomazzo:2007ti}, {\sc SACRA} \cite{Yamamoto:sacra}, 
the LSU-LIU-BYU-PI collaboration code \cite{Palenzuela:2009yr}, and the Illinois group code \cite{Etienne}.
A number of other groups have also developed, or are currently developing,
working GRHD codes.  As many results of astrophysical interest do not require including electromagnetism (EM), 
these codes are also an exciting development.
In addition to implementing several stable evolution schemes that satisfy the constraints of Einstein's equations along with other
consistency requirements (such as preserving the divergence-free character of the magnetic field for GRMHD), most of the aforementioned codes 
have either incorporated, or are currently developing, more relevant physics for problems involving white dwarfs, neutron stars, and any gaseous
environment, such as photon and neutrino transport, and stellar equations-of-state (EOS), ranging from simple polytropes to microphysical EOSs.
While most of the work in this field has so far centered on mergers of binaries that include a neutron star or white dwarf, and is therefore
beyond the scope of this review, there have been a few results from GR(M)HD codes applied to BHBs in gaseous environments.

The Illinois group estimated the accretion luminosity from a BHB immersed is a gaseous environment using GRHD, and compared the result with
the luminosity of a single black hole of the same total mass, immersed in the same environment \cite{Ill1}.  By estimating the contributions
from bremsstrahlung and synchrotron emission, they predicted an enhancement of luminosity by three orders of magnitude at BHB merger, compared to the
single-black-hole case.  If true, this result would have a significant impact on the search for EM counterparts of GW
signals from BHBs.

In addition to accretion luminosity from a surrounding disk, significant luminosity may accompany BHB mergers in the form of jets,
particularly in the presence of an EM field.  Much progress has been made in studying this possibility.  In \cite{Mosta:2009rr},
which built upon earlier studies \cite{Palenzuela:2009yr,Palenzuela:2009hx}, the Einstein-Maxwell equations were evolved for the case of a BHB
immersed in an EM field, with different BHB spin configurations.  By calculating the Poynting flux, they showed that the direct
EM emission was 13 orders of magnitude smaller than the GW emission.  Furthermore, as the EM frequency evolution tracked that for
GWs, the direct EM signature would occur in the mHz range for supermassive BHBs, and would therefore be at far too low a frequency
to be detectable.

However, by combining the evolution of the Einstein-Maxwell equations with a tenuous plasma, which is evolved using the force-free approximation
\cite{Blandford:1977ds,pulsar} wherein the inertia of the plasma is neglected, the authors of \cite{Palenzuela:2010nf,Palenzuela:mag} 
were able to calculate the synchrotron radiation resulting
from the acceleration of the plasma due to the EM Poynting flux previously discussed, upconverting that energy to GHz, and thereby making
it detectable by X-ray observatories.  The radiation transitions from an $m=2$ multipolar structure to an $m=0$ structure, which
has implications for determining whether the EM emission is in fact of the same origin as a given GW observation, since there will likely
be many variable EM sources within the error ellipse of any GW observation.  However, the most interesting aspect of \cite{Palenzuela:2010nf},
from a theoretical standpoint, is the fact that the EM emission is generated by a BHB consisting of two nonspinning black holes.  Since
the emission from a single black hole is most typically associated with the Blandford-Znajek (BZ) mechanism \cite{Blandford:1977ds}, which
requires that the black hole be spinning in order to operate, this new result may be surprising at first.  The flux in the BZ process originates
from the fact that a spinning black hole will twist magnetic field lines, which are assumed to be anchored to a distant accretion disk.
This twisting gives rise to the EM Poynting flux, with the torquing of the magnetic field lines causing a
decrease in the black hole's spin. 

In the case of a BHB, it has now been shown that there is a significant contribution to EM emission that is a generic feature, depending
only on the presence of an orbiting binary and an EM field.  Indeed, given the observation of protostellar jets and their
apparent similarities to their extragalactic counterparts \cite{CarrascoGonzalez:2010da}, it is possible that the conditions for jet formation
may be far more generic than was previously thought, with the presence of plasma and an appropriate magnetic field strength and
geometry being the only clear prerequisites.  

In the context of BHBs, the existence of jets, independent of spin, may not be as surprising
when thought of within the framework of the membrane paradigm \cite{MembraneParadigm}, wherein the black hole event horizon is reinterpreted
as a two-dimensional viscous membrane, and the standard Maxwell equations in three dimensions govern the dynamics of EM
fields evolving along a sequence of hypersurfaces.  It is well understood that, from this viewpoint, the BZ effect can be calculated
using simple circuit equations.  In this picture, a magnetic field threading the black hole and frozen in to a distant accretion disk serves the role
of electrical wires, with charges spiraling along them.  The disk and the horizon membrane serve as resistors, and the potential results from
the twisting of the magnetic field lines by the spin of the black hole.  

The authors of \cite{Palenzuela:2010nf,Palenzuela:mag} suggested that, in the case of a BHB, the Faraday induction results from the orbital
motion of the two black holes through the fixed magnetic field, and could also be understood through application of the membrane paradigm.
We carry out the calculation here, which does indeed bear out the scaling behavior suggested in \cite{Palenzuela:2010nf,Palenzuela:mag}.
Assuming that the plasma is low-density and the force-free approximation applies,
that the orbit is in the \{$\rho$, $\phi$\} plane with orbital velocity $v$ (though any relative velocity between the
black hole and the magnetic field will have the same effect),
and that the background magnetic field is given by $\vec{B}=B\hat{z}$ (in cylindrical coordinates \{$\rho$, $\phi$, $z$\}), the induced potential
from one orbiting hole is then given by
\beq
V=\oint \alpha d\vec{l}\cdot\vec{E} = -\oint \alpha d\vec{l}\cdot(\vec{v}\times \vec{B}) = -2 \alpha r_{\rm H}vB \,
\label{eq:pot}
\eeq
where $\alpha \equiv \sqrt{1-2M/r}$ is the lapse between hypersurfaces of constant Schwarzschild time, $M$ the combined rest mass
of both black holes, and $r_{\rm H}=M$ is the horizon radius of either hole.  The contour integration for employing Faraday's law in Eq.~\ref{eq:pot}
is somewhat subtle.  We place one side along one hemisphere of the stretched horizon, and assume the opposing side resides far away, 
where the $\vec{B}$-field is weak, so that its contribution to the integral vanishes.
We note that this region is assumed to be far away from the computational
domain in \cite{Palenzuela:2010nf,Palenzuela:mag}, where the $\vec{B}$-field is uniform.  The other sides of the contour are
perpendicular to $\vec{v}\times \vec{B}$ where it remains uniform, so that those contributions also vanish, and only the integral
along the horizon remains.

\begin{figure}
\includegraphics[trim = 0mm 0mm 0mm 0mm, clip, scale=.16, angle=0]{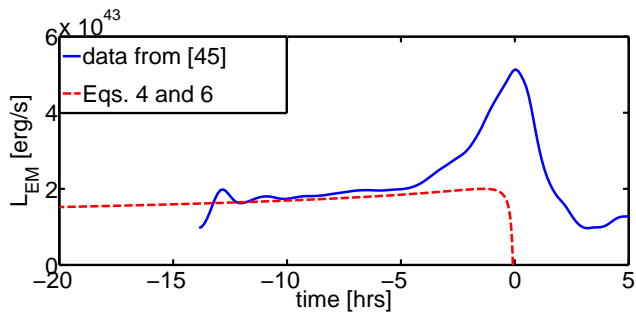}
\caption
{Comparison of the numerical result from Fig.~4 of \cite{Palenzuela:2010nf} (solid line) for the Poynting flux from two orbiting nonspinning black holes
with the simple model given by Eqs.~\ref{eq:pow} and \ref{eq:roft} (dashed line).}
\label{fig:lum}
\end{figure}

To leading order, the orbital velocity is given by $v=\sqrt{M/r}$, so that Eq.~\ref{eq:pot} can be re-expressed as
\beq
V=-2 BM\sqrt{\frac{M}{r}\left(1-\frac{2M}{r}\right)}\,.
\label{eq:pot2}
\eeq
Using the remarkable result that the membrane has an effective resistance, $R_{\rm H}$, of 377 $\Omega$ (or 4$\pi$ in geometrized units)
\cite{Damour_1979}, 
and bearing in mind that each hole emits in both hemispheres, we can solve for the total Poynting flux from both
orbiting holes:
\beq
L=\frac{V^2}{R_{\rm H}}=\frac{4}{\pi}B^2M^2\frac{M}{r}\left(1-\frac{2M}{r}\right)\,.
\label{eq:pow}
\eeq
We note that it is often assumed that the astrophysical load also contributes a resistance, $R_L \approx R_H$ \cite{MembraneParadigm,Palenzuela:mag},
but as this is rather \emph{ad hoc} and depends on poorly understood astrophysics, 
and there is no apparent way that a distant astrophysical load could manifest itself in the simulations
of \cite{Palenzuela:2010nf,Palenzuela:mag}, we neglect this contribution.
Eq.~\ref{eq:pow} predicts a peak luminosity at $r=4M$, which we can express in cgs units with appropriate scaling relationships as
\bea
L_{\rm max} &=& \frac{1}{2\pi}B^2M^2= \frac{4}{\pi a^2} L_{\rm BZ} \nonumber \\
&=& 2\times 10^{43}\, {\rm erg/s} \left(\frac{B}{10^4\,{\rm G}}\right)^2\left(\frac{M}{10^8\,\Msun}\right)^2\,,
\label{eq:powmax}
\eea
where $a$ is the dimensionless spin parameter of the Kerr black hole for the single black hole BZ process ($0\leq a \leq 1$), 
and $L_{\rm BZ}$ is understood
to be the luminosity from the BZ process for a single black hole of spin $a$ and mass $M$ (to approximate the mass of the merged remnant).
We emphasize that this estimate is only valid prior to merger, and indeed predicts a vanishing luminosity for $r=2M$, so that the merger and
its accompanying burst of luminosity seen in \cite{Palenzuela:2010nf,Palenzuela:mag} are not included.  However, if we combine Eq.~\ref{eq:pow}
with the leading order relationship for $r(t)$, 
\beq
r=\frac{5M}{128}(t_c-t)^4\,,
\label{eq:roft}
\eeq
we see excellent agreement between our simple model and the data from Fig.~4 in \cite{Palenzuela:2010nf}
(see Fig.~\ref{fig:lum}).

We also note that this model differs significantly from that suggested in \cite{Lyutikov}.  Using Eq. 7 in \cite{Lyutikov} to find the prediction
for $L_{\rm max}$ in that model, we find that the prediction is approximately four orders of magnitude larger than the numerical result found
in \cite{Palenzuela:2010nf}.  While neither model should
be expected to perform well in predicting the true peak luminosity, which occurs in a highly dynamical and strong-field regime,
the suggestion in \cite{Lyutikov} that $L\propto M^3 B^3$ yields vastly different predictions from Eq.~\ref{eq:pow}, and disagrees with
the scaling suggested in \cite{Palenzuela:2010nf}. 

\subsection{State of the art}

It is useful to summarize the boundaries of parameter space that have currently been explored.  While somewhat arbitrary, a summary
of the extreme cases of different quantities that have been simulated is useful and serves as a snapshot of the current status of the field.
We include a set of physical quantities from NR simulations, the superlative result for that quantity, and the reference
wherein that result was found, in Table \ref{Table:summary}.

\begin{table}
\begin{center}
\begin{tabular}{l|c|c}
\hline \hline
Most waveform cycles                             & 64                     & \cite{Mrouecomm} \\
\hline
Smallest mass ratio                              & 1/100                  & \cite{Lousto:2010ut} \\
\hline
Largest kick (astrophysical)                     & 3254 km/s              & \cite{Dain:2008ck} \\
\hline
Largest kick (hyperbolic)                        & 9589 km/s              & \cite{Healy:2008js} \\
\hline
Largest initial spin ($S_{\rm o}/M_{\rm ADM}^2$) & 0.97                   & \cite{Lovelace:2008tw} \\
\hline
Largest final spin ($S_{\rm f}/M_{\rm ADM}^2$)   & 0.96                   & \cite{Sperhake:2009jz} \\
\hline
Lowest eccentricity                              & $\mathcal{O}(10^{-5})$ & \cite{Mroue:2010re} \\
\hline
Most energetic ($E_{\rm rad}/M_{\rm ADM}$)       & 0.35                   & \cite{Sperhake:2009jz} \\
\hline \hline
\end{tabular}
\end{center}
\caption{Summary of the most extreme BHB systems that have been evolved to-date in numerical relativity simulations, including the
extreme quantity, its value, and the reference for that simulation.
\label{Table:summary}}
\end{table}

The fact that a number of the extremes have not occurred within the past year is perhaps indicative that the tide of NR
breakthroughs that began in 2005 has begun to ebb.  However, as the pace of 2005-2009 could not be maintained indefinitely,
it is our hope that the slowdown is simply the start of a period of more modest, but sustainable, growth in the field.  We emphasize
that the list of superlatives is restricted to results that have been achieved in full BHB simulations, as opposed
to results from single-black-hole simulations, or extrapolations of a set of BHB simulations to some extreme.  We also emphasize that
we are summarizing the extremes in physical systems that have been simulated, and not including extremes in the numerics (such as
smallest phase error, etc.), as any set of such results is arbitrary, and difficult to summarize in an unambiguous way.

\section{Theoretical Advances}
\label{sec:theory}

Apart from achievements in simulating new systems, a burgeoning subfield of NR is the ongoing effort
to draw general theoretical insights from the ever-growing body of available NR simulations.
Independent research projects at Caltech \cite{Keppel,Lovelace:2009dg} 
and Chicago \cite{Gralla} have endeavored to gain analytical insight into the nonlinear dynamics
of momentum transfer in merging BHBs, particularly with respect to the bobbing and ultimate recoil.  Substantial work has been done
to condense the body of simulations into a simple fitting formula, based on either the final state of the binary just prior
to merger \cite{Lousto:2009ka} or on the initial conditions at wider separations \cite{vanMeter:2010md}.  More general
efforts to understand the complete final state of the merged remnant, including the final mass and spin, have been successful
in predicting subsequent numerical results \cite{Rezzolla:2008sd,Lousto:2009mf}.

Progress has also been made in moving beyond purely phenomenological fits of merger waveforms.  Previously, progress was made by fitting
to natural extensions of the post-Newtonian \cite{Ajith:2009bn} or effective-one-body formalisms \cite{Buonanno:2007pf,Damour:2009kr}.
More recently, attempts have been made to apply our improved theoretical understanding of the dominant dynamics during merger in order
to formulate more physically-motivated merger waveforms.  These attempts are an effort to move beyond simply bridging the gap between the well-understood
inspiral and the well-understood ringdown, in order to encapsulate the key features of generic mergers, in the hopes that the results will
prove to be predictive for as-yet-unsimulated (or currently unsimulatable) systems.  In \cite{Tiec:2009yf,Tiec:2009yg}, the close-limit
approximation was applied in an effort to predict the recoil of comparable-mass BHBs.  This research program has a long history, but the
novelty in these investigations is the quality and analyticity of the post-Newtonian initial data, and the existence of full
NR results, which largely validate the effectiveness of this procedure.  Similarly, in \cite{Nichols}, a hybrid of
post-Newtonian and black hole perturbation theory techniques was used to study the radiated energy and momentum and the gravitational
waveforms for head-on collisions.  The results compared favorably with full NR simulations.  

In \cite{Baker:2008mj,McWilliams:2010eq},
the ``implicit rotating source'' model for merger waveforms was developed, and its results were rigorously compared with full
numerical simulations for nonspinning systems with mass ratios $q\leq 1/6$.  This model was built on the observation that merger
waveforms from numerical simulations behave as though they were generated by a rigid rotator.  Specifically, the frequency evolution
of all $\ell=m$ modes, when weighted by $m$, are equal across all modes and mass ratios, consistent with multipoles of
a single object rotating with a well-defined frequency.  The angular momentum at merger is also proportional to the frequency, as one
would expect for a rotator with a well-defined moment of inertia.  These and other indicators led the authors of \cite{Baker:2008mj}
to formulate a physically-motivated merger waveform, which naturally ties to the final ringdown phase of the waveform, whereas
most other methods tie to the late inspiral waveform, at a point where it may have already accumulated a large degree of error.
Combined with the aforementioned models for the final mass and spin of the merged remnant across parameter space, this model may
provide a novel method for predicting the merger waveform for generic systems, with a minimum of fine-tuning.  

The commonality of
all the approaches to gaining analytical insight into merger dynamics is the observation that the merger, though happening in the
strong-field, nonlinear regime, seems to be well-described by tools developed primarily for linear perturbative analyses in the weak-field
regime.  This observation is one of the most fundamental results that has been made possible by numerical relativity.

\section{Conclusions}
\label{sec:conc}

We have presented a brief overview of the current state of
black-hole binary simulations in numerical relativity.  While progress has slowed somewhat since the ``gold rush'' period following the breakthroughs
in 2005--2006, there have still been a number of novel results of significant astrophysical and theoretical interest.
In the past year, progress has been made in simulating longer waveforms overall, as well as systems with more extreme mass ratios
and larger spins.  Advances in methodology have significantly enhanced the accuracy and efficiency achievable by state-of-the-art codes.
Progress has also been made in using the body of available numerical simulations to inform a greater theoretical understanding
of the strong-field behavior of black-hole binaries.  The inclusion of matter and electromagnetic fields in 
black-hole binary simulations has led to unexpected
discoveries, which will have significant astrophysical implications in the years to come.  The field of numerical relativity
has now entered a period of more modest, but hopefully sustainable growth, with a substantial amount of discovery space that remains
to be explored.

\begin{acknowledgments}
We wish to thank Luis Lehner for useful discussions of the dual BHB jet result and for providing his numerical data, 
and Bernard Kelly and Jennifer Seiler for helpful feedback on the manuscript.
\end{acknowledgments}

\end{document}